\documentclass{aa}  

\usepackage{natbib}
\usepackage{graphicx}
\usepackage{txfonts}
\newcommand{\be}{\begin{equation}}
\newcommand{\ee}{\end{equation}}
\newcommand{\ba}{\begin{eqnarray}}
\newcommand{\ea}{\end{eqnarray}}

\begin{document}


\title{3D propagation of relativistic solar protons\\ through interplanetary space}

\titlerunning{Relativistic solar proton propagation}

   \author{S.~Dalla\inst{1}
          \and
           G.A.~de Nolfo\inst{2}
          \and
           A.~Bruno\inst{2}
          \and
           J.~Giacalone\inst{3}
          \and 
           T.~Laitinen\inst{1} 
          \and 
          S.~Thomas\inst{1} 
          \and 
          M.~Battarbee\inst{4}
           \and 
          M.S.~Marsh\inst{5}         }

   \institute{Jeremiah Horrocks Institute, University of Central Lancashire,
    Preston, PR1 2HE, UK \\
              \email{sdalla@uclan.ac.uk}
         \and
            NASA Goddard Space Flight Center, USA \\
         \and
            University of Arizona, USA \\
         \and
            University of Helsinki, Finland            
         \and
            Met Office, Exeter, UK           
             }

   \date{Received January 2019; }

% \abstract{}{}{}{}{} 
% 5 {} token are mandatory
 
  \abstract
   {Solar Energetic Particles (SEPs) with energy in the GeV range can propagate to Earth from their acceleration region near the Sun and produce Ground Level Enhancements (GLEs). The traditional approach to interpreting and modelling GLE observations assumes particle propagation only parallel to the magnetic field lines of interplanetary space, i.e. it is spatially 1D. Recent measurements by PAMELA have characterised SEP properties at 1 AU for the $\sim$100 MeV--1 GeV range at high spectral resolution.}
   {We model the transport of GLE-energy solar protons using a 3D approach, to assess the effect of the Heliospheric Current Sheet (HCS) and drifts associated to the gradient and curvature of the Parker spiral. We derive 1 AU observables and compare the simulation results with data from PAMELA.}
   {We use a 3D test particle model including a HCS. Monoenergetic populations are studied first to obtain a qualitative picture of propagation patterns and numbers of crossings of the 1 AU sphere. Simulations for power law injection are used to derive intensity profiles and fluence spectra at 1 AU. A simulation for a specific event, GLE 71, is used to compare with PAMELA data. } 
   {Spatial patterns of 1 AU crossings and the average number of crossings are strongly influenced by 3D effects, with significant differences between periods of $A^+$ and $A^-$ polarities. The decay time constant of 1 AU intensity profiles varies depending on the position of the observer and is not a simple function of the mean free path as in 1D models. Energy dependent leakage from the injection flux tube is particularly important for GLE energy particles, resulting in a rollover in the spectrum.  }
   {}


   \maketitle
%
%
%-------------------------------------------------------------------

\section{Introduction}\label{sect_intro}

Ions of relativistic energies can be accelerated at or near the Sun during flare and Coronal Mass Ejection (CME) events. 
When detected in the interplanetary medium, for example near Earth, they constitute the high energy portion of the spectrum of Solar Energetic Particles (SEPs) \citep{Mew2012,Coh2018}, whose properties are an important tracer of the acceleration processes and of the propagation through the Interplanetary Magnetic Field (IMF) .

Relativistic solar ions may produce secondary particles when they interact with Earth's atmosphere, causing so-called Ground Level Enhancements (GLEs), observed in ground-based neutron monitor data \citep{Bel2010, Nit2012, Gop2012,  McC2012, Mis2018}. Protons in the energy range $\sim$0.5--30 GeV are thought to be the main contributors to GLEs (eg.~\citealt{McC2012}). GLEs are much less frequent than SEP events detected by spacecraft instrumentation, which is typically sensitive to protons up to $\sim$100 MeV. Only 72 GLE events have been detected by the worldwide network of neutron monitors from 1942 to the present time (eg.~\citealt{Bel2010}).

Recent SEP observations from the PAMELA detectors have allowed us to fill the particle energy gap between traditional spacecraft instrumentation and neutron monitors, and routinely detect relativistic solar protons in the range from $\sim$100 MeV to a few GeV \citep{Adr2015,Bru2018}. The new observations call for modelling tools that describe the acceleration and propagation of particles at these energies. In addition, simulations of propagation through the IMF are necessary to relate the detections of high energy SEPs at 1 AU to the numbers of interacting particles at the Sun which produce solar $\gamma$-ray events detected by FERMI \citep{DeN2019,Sha2018, Kle2018}.

A number of studies have modelled the propagation of relativistic solar protons through the IMF using spatially 1D descriptions, to interpret neutron monitor observations. The effect of magnetic field turbulence on particle propagation is typically described as pitch-angle scattering, characterised by a mean free path $\lambda$.
\cite{Bie2004} and \cite{Sai2005} used a model based on the focused transport equation to fit data for two GLE events.  
\cite{Str2017} used a focused transport model to calculate rise and decay times of GLEs. 
\cite{Heb2018} combined 1D propagation within interplanetary space of GLE-energy particles with trajectory integration through magnetospheric configurations. 
\cite{LiL2019} found analytical expressions for the flux profile and anisotropy of relativistic protons using a focused transport approach within specific scattering conditions, and used them to fit the 2005 January 20 GLE.
The 1D approximation, which assumes that particles remain tied to the magnetic field line on which they were injected, is therefore the standard approach used to model the interplanetary propagation of solar relativistic protons, and to analyse GLE observations (e.g.~\citealt{Nit2012}).
Within this approximation, the effects of IMF polarity and of the heliospheric current sheet on the propagation of relativistic protons are neglected. 

%-------------------------------------- Two column figure (place early!)
   \begin{figure*}
   \centering
   \includegraphics[scale=.6]{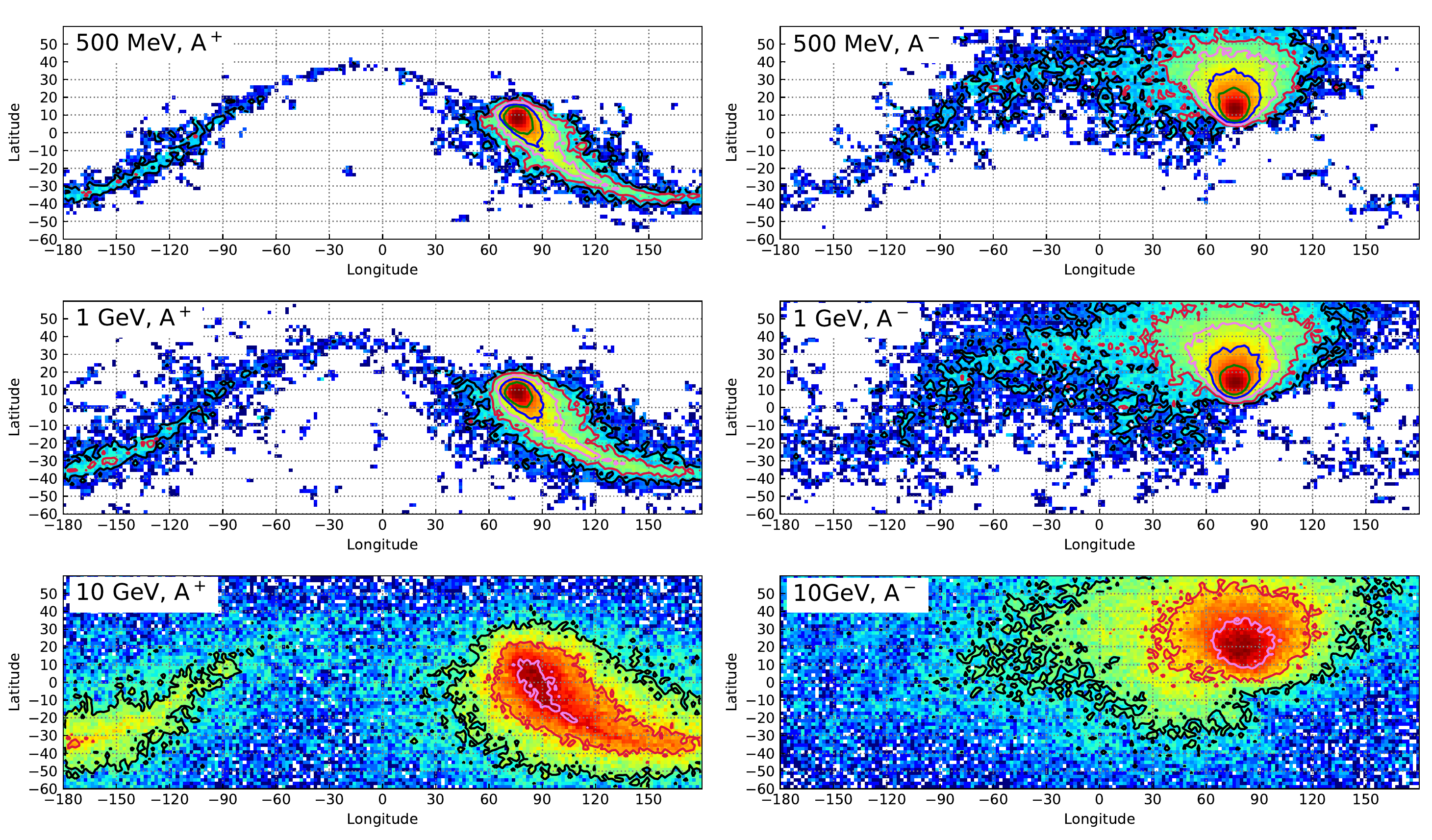}
 \caption{Maps of cumulative 1 AU crossings in a heliocentric coordinate system corotating with the Sun, for monoenergetic SEP proton populations, with energy as indicated in each panel. {\it Left column}: A$^+$ configuration of the IMF; {\it right column}:  A$^-$ configuration. The  8$\times$8$^{\circ}$ injection region at the Sun is located at longitude 76$^{\circ}$ and latitude 11$^{\circ}$, above the HCS, and the zero of the coordinate system on the 1 AU map has been shifted so that the flux tube through the injection region appears at N11W76 on this map. The tilt of the HCS is $\alpha_{nl}$=37$^{\circ}$. All simulations used $N$=10,000 protons, solar wind speed $v_{sw}$=400 km s$^{-1}$ and mean free path $\lambda$=0.1 AU. Contour lines are plotted for the following values of the number of crossings: 1000 ({\it green}), 316 ({\it blue}), 100 ({\it lilac}), 31 ({\it red}) and 10 ({\it black}).}
\label{fig.map_plots}  
     \end{figure*}
%---------------------------------------

 A well developed theory of the propagation of Galactic Cosmic Rays (GCRs), relativistic protons originating outside the heliosphere and propagating through the IMF, has been used to describe GCR modulation over several decades (e.g.~\citealt{Pot2017}). Within GCR models, dealing with e.g.~protons of energies above $\sim$1 GeV, a spatially 3D description of particle propagation is thought to be necessary, due to effects such as IMF gradient and curvature drifts, diffusion in the direction perpendicular to the average field, and the influence of the Heliospheric Current Sheet (HCS) (e.g. \citealt{Par1965,Kot1983,Bur2012}).

It is the aim of this paper to model the interplanetary propagation of relativistic protons by means of a fully 3D approach, allowing us  to discuss the effects of the HCS and IMF polarity on 1 AU observables.
Our earlier work has pointed out that drifts due to the gradient and curvature of the Parker spiral IMF do affect the propagation of SEPs, with their importance increasing with particle energy and being particularly significant for heavy ions \citep{Mar2013,Dal2013, Dal2017a, Dal2017b}. Analysis of the role played by a flat HCS \citep{Bat2017} and by a wavy HCS \citep{Bat2018a} on SEPs injected with power law distributions in the range 10-800 MeV demonstrated the role of injection region location and IMF polarity, and elucidated how the HCS provides an efficient means for particle transport in longitude.

In this paper, we focus on relativistic protons in the energy range from a few hundred MeV to 10 GeV and demonstrate the need for an approach that describes propagation as fully 3D, unlike the traditional approaches to GLE modelling. In particular we show that once a 3D approach is adopted and a HCS is introduced in the model, significant dependencies of 1 AU observables on the magnetic polarity of the IMF are observed. We point out how the latter affects time-intensity profiles and spectra, analysed at multiple locations defined with respect to the magnetic flux tube with nominal connection to the centre of the injection region.
We also focus on a specific relativistic particle event for which PAMELA detected protons over a wide energy range, GLE 71, occurring on May 17, 2012, and compare our modelled observables with preliminary data from its detectors \citep{Adr2015,Bru2018}. This is the first comparison of SEP PAMELA data with a model.

In Section \ref{sec.monoe} we present our model and the results of simple monoenergetic injection simulations, including a discussion of the number of 1 AU crossings. In Section \ref{sec.power} we consider a power-law distribution of relativistic protons and discuss how transport through interplanetary space shapes the 1 AU observables over a grid of locations.
In Section \ref{sec.pamela} a comparison between our model and PAMELA intensity profiles is presented for GLE 71.
We discuss our results in Section \ref{sec.discuss}.

\section{Simulations of monoenergetic populations}\label{sec.monoe}

We model relativistic proton propagation through space by integrating particle trajectories in 3D via a full orbit test particle code \citep{Mar2013, Dal2005}.
Particle acceleration is not modelled and injection characteristics of the accelerated population are specified as input.
The IMF is characterised by two polarities separated by a model wavy HCS \citep{Bat2018a}. Using standard terminology from GCR studies, the configuration with magnetic field pointing outwards in the northern hemisphere and inwards in the south is referred to as $A^+$ and that with opposite polarity as $A^-$.

Scattering due to turbulence in the interplanetary magnetic field is simulated by means of the so-called  \lq ad-hoc scattering\rq\  method. A sequence of Poisson-distributed scattering events for each particle is generated, compatible with a mean scattering time $t_{scat}$=$\lambda/v$, where is $\lambda$ the specified value of the mean free path and $v$ the particle's speed. At each scattering event, the direction of the particle's velocity is reassigned randomly from a uniform spherical distribution \citep{Kel2012, Mar2013}.
This method for describing scattering within SEP test particle simulations has been used by a number of groups over the years (e.g. \citealt{Koc1998, Pei2006, Cho2010, Kel2012, Mar2013}). The value of the scattering mean free path within the ad-hoc scattering method is equivalent to that of traditional diffusion descriptions.
 \citet{Koc1998} directly compared the ad-hoc scattering approach (termed small time-step isotropisation (SSI) model in their work) with a traditional diffusion-convection description: they obtained very close agreement in SEP time intensity profiles at 1 AU when the same value is used for $\lambda$ in the ad-hoc test particle approach and as parallel mean free path in the diffusion-convection model (see Figure 4 of \citealt{Koc1998}).
 They also compared the results of the ad-hoc scattering description, in which the pitch-angle may change by a large angle during a scattering event, with two small angle scattering descriptions within the focussed transport equation, one isotropic and one anisotropic (indicated in their work as IAS and AAS respectively). They found that for the same value of the mean free path, 1 AU time intensity profiles for all these models are very similar, with some differences in the peak intensities and closely matching decay phases and durations (see Figure 5 of \citealt{Koc1998}).

There is no consensus within the literature about the degree of scattering experienced by GLE energy protons in their travel to 1 AU. In the simulations of \cite{Bie2004} and \cite{Sai2005}, fitting to GLE data, within their 1D model, yielded $\lambda$ $\sim$ 0.1 AU. \cite{LiL2019} were able to reproduce observations only by assuming different scattering conditions for different phases of a GLE event: at the beginning of the event they used $\lambda$=4 AU, meaning near scatter-free conditions, while later in the event strong scattering, with $\lambda$ an order of magnitude smaller, was required to fit the data.
In our simulations we consider a variety of mean free paths, kept constant over time and we neglect the dependence of $\lambda$ on energy for the relativistic particle energy range we consider.

In this initial study we do not explicitly introduce a term describing perpendicular transport associated with turbulence in the solar wind magnetic field, for example due to magnetic field line meandering \citep{Lai2016}. Our scattering description does implicitly result in minor random-walk of the particle's gyrocenter across the magnetic field, of the order of a Larmor radius, at each scattering event. This finite Larmor radius effect is small and it is negligible compared to typical cross-field diffusion due to random-walk, or meandering, of turbulent magnetic field lines \citep[e.g.][]{Jokipii1966,  GiaJok1999, Fra2011}. Thus turbulence-associated perpendicular transport is not included in our simulations and motion across the magnetic field seen in our results is mainly due to drift and HCS effects.

It is instructive to analyse the propagation of monoenergetic populations of relativistic protons, to visualise how 1 AU observables vary with particle energy. 
Each monoenergetic population consists of $N$=10,000 particles, injected instantaneously from a small region at the Sun of angular extent 8$\times$8$^{\circ}$, located at $r$=2 $R_{sun}$. While in actual SEP events the source region may in fact be much more extended, the key properties of the propagation are revealed more clearly if the injection is localised within the model.

The magnetic field in the simulation is given by a Parker spiral field. We use the method described by \cite{Bat2018a} to include a HCS.
When the presence of a HCS is taken into account, parameters of the HCS such as the tilt angle $\alpha_{nl}$, the polarity of the IMF and the position of the injection region with respect to the HCS, have a strong influence on the particle propagation \citep{Bat2018a} .

\subsection{Maps of 1 AU crossings} \label{sec.maps}

Figure \ref{fig.map_plots} shows longitude-latitude maps of crossings of the 1 AU sphere summed over the entire duration of the simulations, for monoenergetic populations at 500 MeV, 1 GeV and 10 GeV, where these populations were followed up to a time $t_f$=61 hr. 
The mean free path $\lambda$ is 0.1 AU. The injection region, corresponding to the dark red pixels, e.g. in the top left plot, is located above the HCS, at longitude 76$^{\circ}$ and latitude 11$^{\circ}$.  The tilt of the HCS is $\alpha_{nl}$=37$^{\circ}$.
The maps show 1 AU crossings in a heliographic coordinate system that is corotating with the Sun. 

The left panels show maps for an $A^+$ configuration and the right panels for $A^-$, so that in the former case particles tend to move towards the HCS and in the latter away from it, due to  gradient and curvature drift in the Parker spiral IMF \citep{Dal2013,Mar2013,Bat2018a}. This motion towards/away from the HCS follows standard GCR patterns \cite{Jok1977}.
Since gradient and curvature drift effects increase with energy, the 10 GeV particles show the largest transport across the field, and for the latter population the peak counts location is southwards of the injection region for the $A^+$ configuration and northwards for $A^-$.
In addition to gradient and curvature drift, HCS drift also affects the spatial patterns in Figure \ref{fig.map_plots}.
In the $A^+$ case, as they reach the HCS by drifting southwards, particles experience a strong westward HCS drift that spreads them efficiently in longitude. In the $A^-$ case, a drift along the HCS is also observed, though it is less pronounced compared to the $A^+$ situation, because gradient and curvature drift tend to move particles away from the HCS, and it is in the opposite direction (eastwards). 

Looking at the bottom panels, for 10 GeV, one can see that although the injection region is only 8$^{\circ}$$\times$8$^{\circ}$ in extent, the entire 1 AU sphere is accessible to particles, despite the fact that the injection was localised. It is interesting to note that at these energies, although rapid transport across the field allows particle access to regions far away from the injection region, it also quickly dilutes the population, making it more difficult for it to be detected above the GCR background. Looking at the two bottom rows, it is clear that over the energy range of interest for GLEs, interplanetary propagation is fully 3D.

The patterns seen in Figure \ref{fig.map_plots} present some differences and similarities to the maps of 1 AU crossings presented by \cite{Bat2018a}: in the latter study, a power law proton population in the energy range 10--800 MeV was considered. Their maps were therefore dominated by $\sim$10 MeV particles, which experience much smaller drift compared to relativistic protons, resulting in a less pronounced drift along the HCS for starting locations that were not directly located on the HCS itself. 
The overall qualitative dependence of patterns on $A^+$ versus  $A^-$ is the same as in \cite{Bat2018a}.

The panels of Figure \ref{fig.map_plots} do not include the effect of corotation, i.e.~the fact that, in the spacecraft frame, magnetic flux tubes filled with particles cross a number of heliospheric longitudes over time. Corotation increases the spatial extent of the event in longitude (for an example of maps with and without the inclusion of corotation see Figure 1 of \citealt{Bat2018a}), however at the energies considered Figure \ref{fig.map_plots} the effects of corotation are less evident than at lower energy.

\subsection{Average number of 1 AU crossings per particle}

%-------------------------------------- Two column figure (place early!)

   \begin{figure*}
   \centering
   \includegraphics[scale=.55]{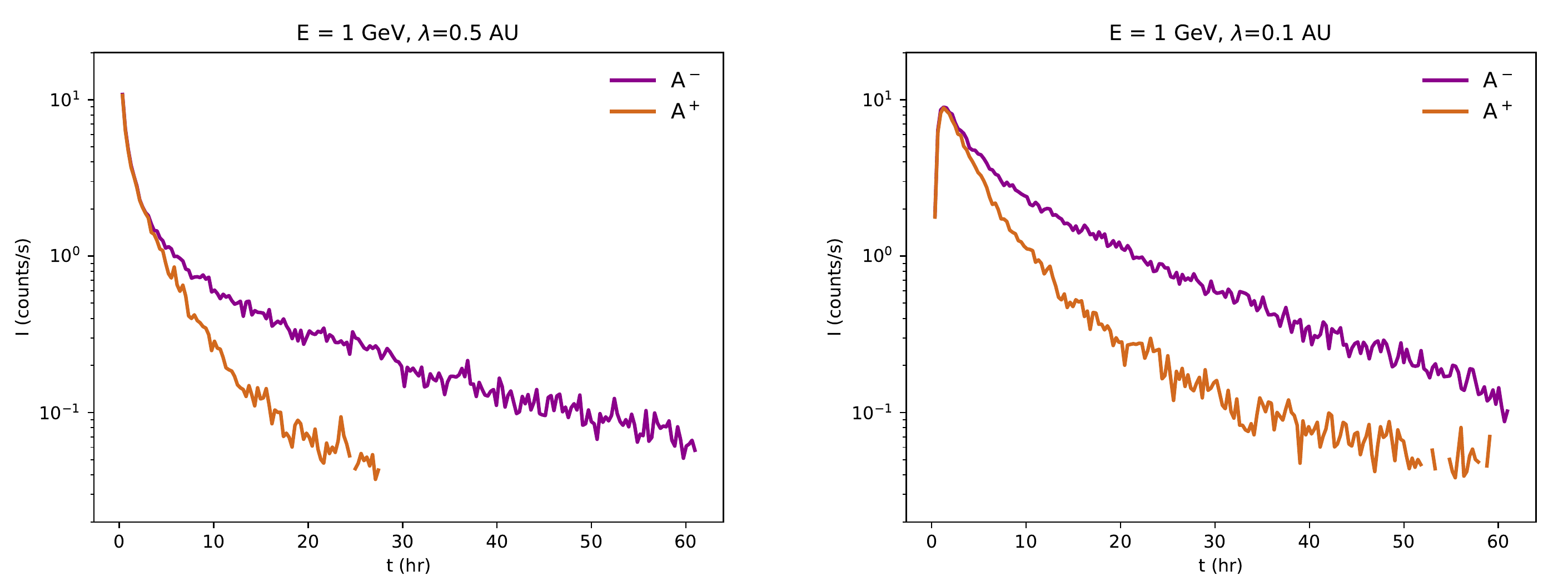}
 \caption{1 AU crossing count rates versus time summed over all heliographic longitudes and latitudes, for a monoenergetic proton population of initial kinetic energy $E$=1 GeV and mean free path $\lambda$=0.5 AU  ({\it left}) and $\lambda$=0.1 AU ({\it right}), for A$^+$ and A$^-$ configurations of the IMF. Other parameters of the simulations are as in Figure 1.}
\label{fig.intens_Ap_Am}  
     \end{figure*}

%

%--------------------------------------------------- One column table
   \begin{table}
      \caption[]{Average number of 1 AU crossings per particle, ${\overline{N}}_{cross}$, as a function of SEP proton kinetic energy $E$, for $A^+$ and $A^-$ configurations, and mean free path $\lambda$.}
      \label{table_ncross}
      \centering
         \begin{tabular}{cccc}
            \hline
            \noalign{\smallskip}
   $\lambda$ (AU)   &   E (GeV)     &  $A^+$  &  $A^-$ \\
            \noalign{\smallskip}
            \hline
            \noalign{\smallskip}
       0.1 &    0.5 &   21 & 30 \\
       0.1 &    1  &     17 &   29     \\
       0.1 &    10  &  14 &  21 \\
       0.5 &    1  &    7 & 11 \\
               \noalign{\smallskip}
            \hline
         \end{tabular} 
   \end{table}
%

   %-------------------------------------- Two column figure - intensity profiles
   \begin{figure*}
   \centering
   \includegraphics[scale=.75]{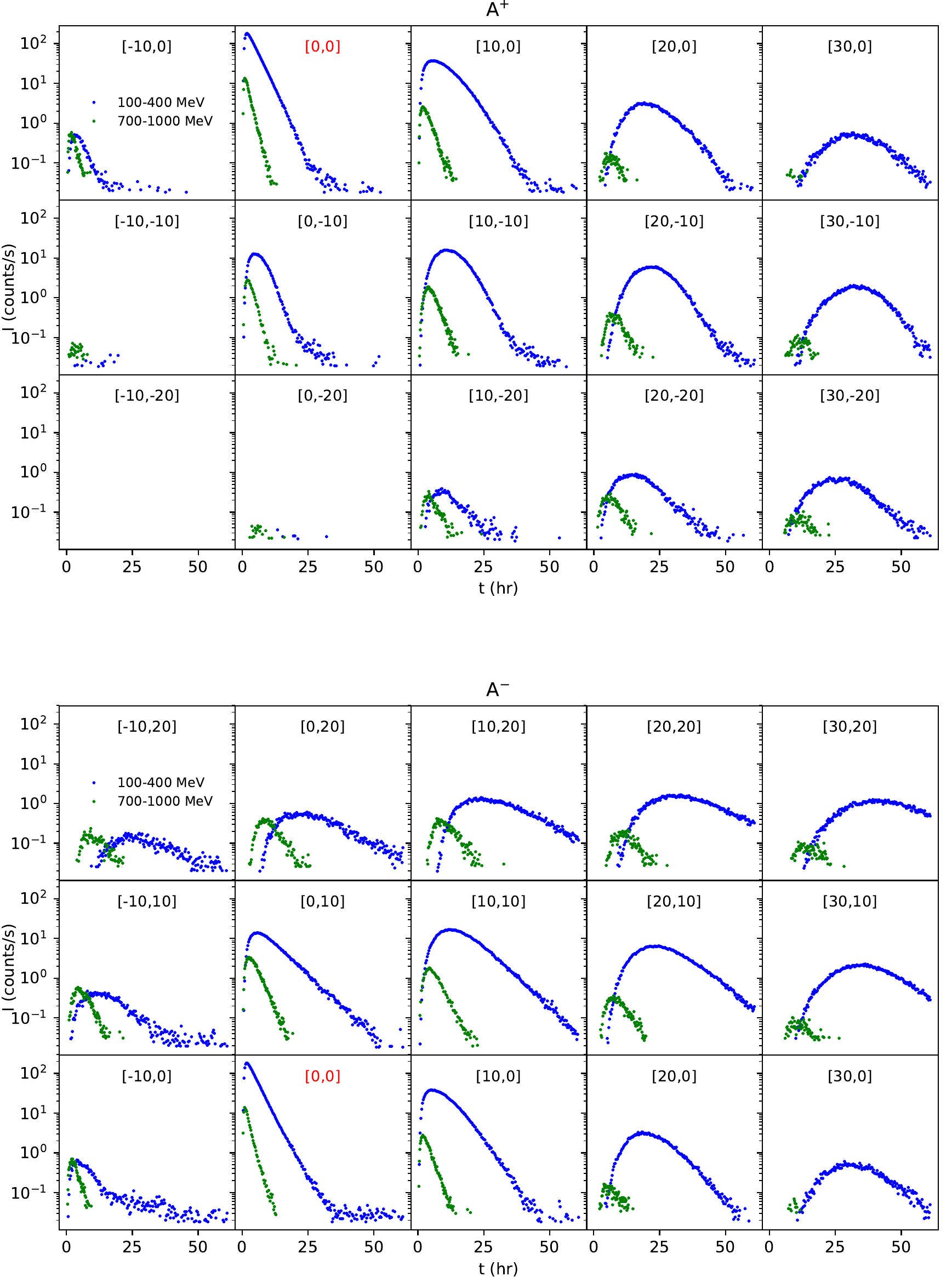}
   \caption{Proton count rates versus time for $A^+$ ({\it top}) and $A^-$ ({\it bottom}) configurations of the IMF, at a variety of 1 AU locations with respect to the best connected location ([0,0]), for a power law population, for the proton energy ranges 100--400 MeV ({\it blue}) and 700--1000 MeV ({\it green}). }
\label{fig.intens_prof} 
   \end{figure*}

 %-------------------------------------- Two column figure - spectra
   \begin{figure*}
   \centering
   \includegraphics[scale=.75]{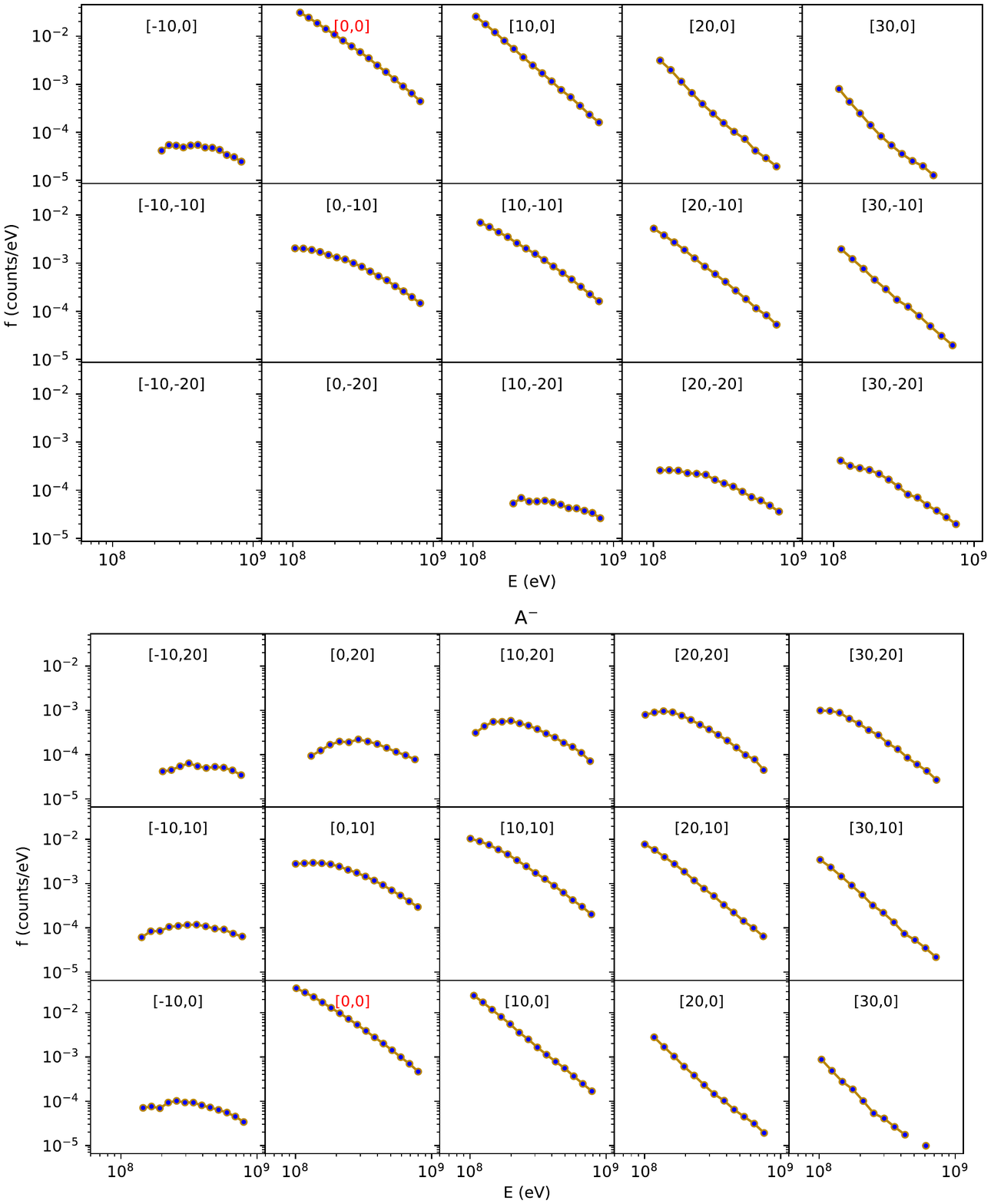}
   \caption{Fluence energy spectra for $A^+$ ({\it top}) and $A^-$ ({\it bottom}) configurations of the IMF, at a variety of 1 AU locations with respect to the best connected location ([0,0]), for a power law population. The solid lines in the [0,0] panels give the slope of the injection spectrum. Parameters of the simulations are as in Figure 3.}
\label{fig.spectra} 
     \end{figure*}

In addition to the spatial patterns of crossings of the 1 AU sphere, it is interesting to consider  ${\overline{N}}_{cross}$, the average number of 1 AU crossings per particle, for a specified SEP kinetic energy.  Particles may cross 1 AU more than once as they scatter back and forth in their propagation, so that this parameter is a strong function of the mean free path \citep{Cho2010}.
${\overline{N}}_{cross}$ is needed to estimate the total number of SEPs at 1 AU from spacecraft detections of fluxes \citep{Mew2008}. 
Therefore knowledge of ${\overline{N}}_{cross}$, for example from transport simulations, allows one to compare 1 AU SEP numbers with the number of interacting particles at the Sun, deduced e.g.~from $\gamma$-ray observations \citep{DeN2019}. 

We derive ${\overline{N}}_{cross}(E)$ from our model, where $E$ is particle energy,  by obtaining the number of 1 AU crossings per particle for each integrated trajectory in our monoenergetic population simulation, with crossings collected over the entire 1 AU sphere and for the duration of the simulation, and calculating its average over the population.
It should be noted that particles do decelerate as they propagate through interplanetary space (see e.g.~\citealt{Dal2015}), however the effect is less prominent at the energies considered here, so that it is a reasonable assumption to take the initial energy as $E$.

Table \ref{table_ncross} displays ${\overline{N}}_{cross}$ values for the $\lambda$=0.1 AU simulations displayed in Figure \ref{fig.map_plots} and, for comparison, a case with  $\lambda$=0.5 AU (see also \citealt{DeN2019}). 
A strong dependence of ${\overline{N}}_{cross}$ on the IMF polarity is therefore deduced from our simulations, with the number of crossings being much larger for $A^-$ polarity than for $A^+$. This behaviour is equivalent to the polarity dependence of fluence that was discussed by  \cite{Bat2018a}.
It should be noted that the distribution of the number of crossings per particle is generally quite broad, so that the standard deviation for the averages in Table \ref{table_ncross} is almost as large as the values themselves.
%For the $\lambda$=0.1 AU simulations displayed in Figure \ref{fig.map_plots}, ${\overline{N}}_{cross}$= 21, 17 and 14 for $E$=500 MeV, 1 GeV and 10 GeV respectively for $A^+$,  while for $A^-$ we obtain ${\overline{N}}_{cross}$= 30, 29 and 21 for the same values of $E$ (see also \cite{DeN2019}). 

Figure \ref{fig.intens_Ap_Am} shows the time evolution of the count rate $I$ (number of 1 AU detections divided by accumulation time), using the whole 1 AU sphere as collection area, for injection energy $E$=1 GeV and for the two polarities. The right hand panel ($\lambda$=0.1 AU) corresponds to the same simulations displayed in the central row of Figure \ref{fig.map_plots}, while the left hand panel has $\lambda$=0.5 AU.
There is a large difference in the time evolution of the count rate depending on the polarity of the IMF, with the  A$^+$ polarity decay being much faster than for A$^-$. 

The reason for the differences between $A^+$ and $A^-$ in Figure \ref{fig.intens_Ap_Am} and Table \ref{table_ncross}   is that in the former configuration, drift along the HCS is more prominent, so that protons move towards the outer heliosphere faster than for $A^-$ and a significantly lower number of 1 AU crossings occur. The reason why the two curves in Figure \ref{fig.intens_Ap_Am} are very similar at early times is that it takes a finite amount of time for particles to drift down to the HCS in the A$^+$  case, at which point HCS drifts set in. Our findings on the influence of IMF polarity on number of crossings per particle is confirmed by a completely independent test particle simulation code with flat HCS \citep{Cho2010, DeN2019}. We note that changing the parameters of the HCS (for example the tilt angle) does not affect ${\overline{N}}_{cross}$ strongly and that its energy dependence (fewer crossings at higher energies) is a result of the particle populations at high energies propagating faster towards the outer heliosphere.

\section{Simulations of power-law populations}\label{sec.power}

We consider a proton population injected with a distribution of energies that follows a power law,
and propagate it through interplanetary space using the same HCS configuration as in Section \ref{sec.monoe}.
We choose a spectral index at injection $\gamma$=2 for the energy range 100 MeV--1 GeV. 
The population is injected from the same location as the monoenergetic runs shown in Figure \ref{fig.map_plots} and with the same parameters. Therefore also in this analysis, we use a small 8$\times$8$^{\circ}$ injection region.

\subsection{Intensity profiles}

To produce intensity profiles, counts are collected over 10$^{\circ}$$\times$10$^{\circ}$ portions of the 1 AU sphere that mimic a variety of observer locations with respect to the injection region. Here the observer is not corotating with the Sun but is in the so-called spacecraft frame.
Figure \ref{fig.intens_prof} shows intensity profiles at a variety of locations
 for the energy ranges 100--400 MeV ({\it blue}) and 700--1000 MeV ({\it green}).
The top grid refers to an $A^+$ IMF configuration and the bottom grid to $A^-$. 
Observer locations are specified using labels $[\Delta\phi_{1AU}, \Delta\delta_{1AU}]$, where $\Delta\phi_{1AU}$ is the 
heliographic longitude and $\Delta\delta_{1AU}$ the heliographic latitude of the observer relative to the Parker spiral field line through to the centre of the particle injection region. The panel labelled [0,0] ({\it red label}) corresponds to an observer connected to the centre of the injection region at the time of injection, and the other panels to less well connected observers ({\it black labels}).  In a 1D model, intensities would be zero everywhere apart from the well connected panel, [0,0].

Moving from left to right along a row in Figure \ref{fig.intens_prof} one can see count rate profiles for observers at the same latitude and progressively more western longitudes (i.e.~source region becoming more eastern). 
Here one can see the important effect of corotation, in the lower energy range, resulting in a less sharp rise phase and later time of peak intensity as the source region becomes more eastern, as already noted in our previous studies \citep{Mar2015, Dal2017a,Lai2018}.
Different rows correspond to different observer latitudes, becoming more southern as one moves downwards.
The observer locations for $A^+$ ($A^-$) have been chosen to reflect the fact that, as shown in the maps of Figure \ref{fig.map_plots}, the spatial extent is mostly downwards (upwards).

Figure \ref{fig.intens_prof} shows that the event duration is much shorter in the 700--1000 MeV range compared to the 100--400 MeV range. This is due to the combination of two effects: 1) the higher energy protons travel away from the inner heliosphere faster and 2) they experience stronger transport across the field due to drift effects (as shown in Figure \ref{fig.map_plots}), resulting in much faster dilution of the population. Therefore more efficient drift across the field does not necessarily mean a higher probability of detection at far away locations, since dilution works against detection above background at a given spacecraft. 
At lower energies, particles are confined inside a \lq cloud\rq\ around the injection flux tube and as a result of corotation they can produce significant count rates over extended times. 

Comparing the top and bottom sets of grids in Figure \ref{fig.intens_prof}, two main differences between $A^+$  and $A^-$ are observed:  the overall spatial extent of the event is larger for the $A^-$ case, in agreement with the monoenergetic 1 AU maps shown in Figure \ref{fig.map_plots}, and at many observers the decay phase tends to last longer in the $A^-$ configuration compared to $A^+$, replicating the behaviour seen for the global crossings in Figure \ref{fig.intens_Ap_Am}.

The slope of the decay phase varies significantly for different locations for a given polarity configuration, as well as between $A^+$  and $A^-$.
Thus in 3D this parameter is not simply a reflection of the value of the mean free path $\lambda$, as would be the case in a 1D model, but it is the result of a number of processes that include IMF polarity and HCS effects and dilution due to transport across the magnetic field. 

\subsection{Fluence spectra}

The fluence spectra for the same locations and configurations as in Figure \ref{fig.intens_prof} are presented in Figure \ref{fig.spectra}.
Although the injection spectrum is a power law with $\gamma$=2, it is evident that a variety of spectral shapes are seen at the different observers, as a result of 3D propagation effects.

The fact that drifts effects are stronger for high energies has an influence on particle spectra: as a result of the dilution effect discussed in Section \ref{sec.maps} at the best connected location the spectrum is no longer a power-law but displays a roll-over. Rollover features are observed in PAMELA spectra \citep{Bru2018}.
At locations away from the well connected ones a variety of features are observed, connected to dilution at high energies and the fact that lower energy particles drift across the field less efficiently. In addition, at the lower end of the energy range shown in Figure \ref{fig.spectra} adiabatic deceleration affects the spectra \citep{Dal2015}.
Our simulation show that any mechanism that produces energy-dependent escape from the flux tube will \lq process\rq\ the injection spectrum.

In addition to the simulation for $\gamma$=2 (as shown in Figure \ref{fig.spectra}) we analysed spectra also for a case with initial injection spectrum with $\gamma$=3 and found that the qualitative behaviour in this case is very similar to that for the case $\gamma$=2, apart from spectra being generally softer.

\section{Comparison with PAMELA data for GLE 71}\label{sec.pamela}

   %-------------------------------------- Two column figure - comparison with PAMELA data - map
   \begin{figure}
   \centering
   \includegraphics[scale=.5]{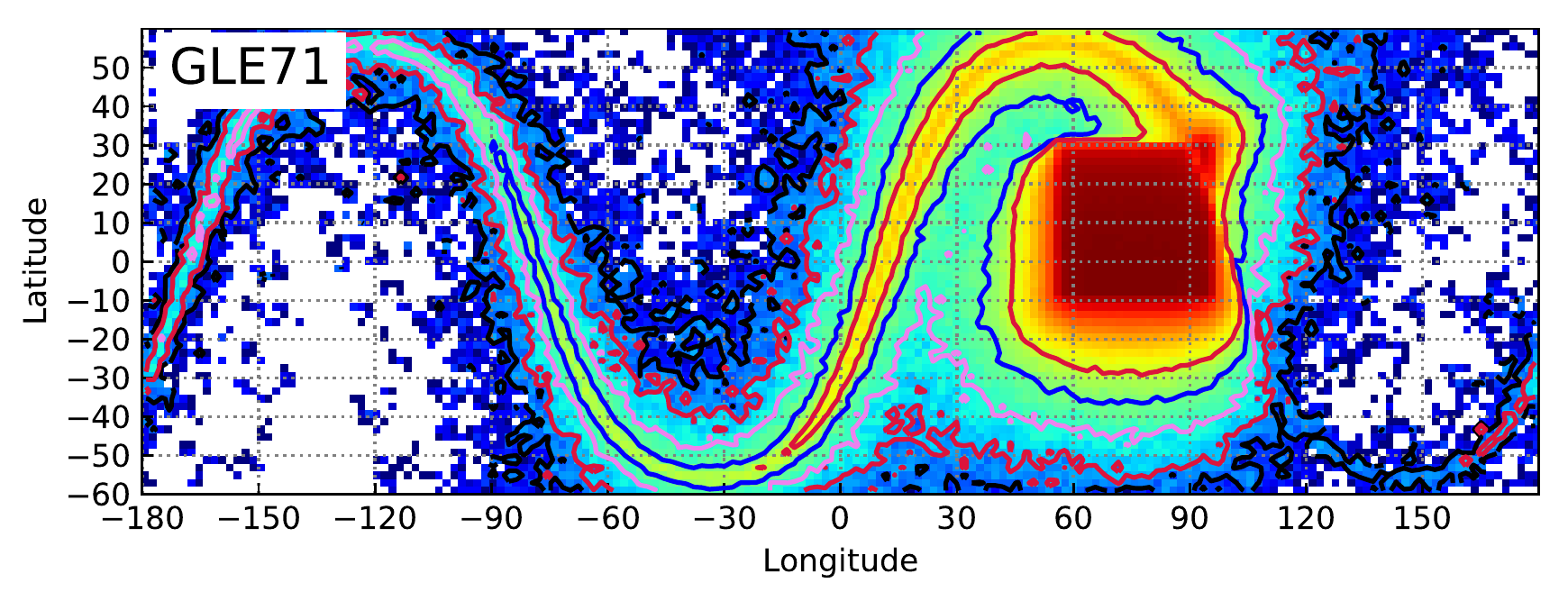}
   \caption{Maps of cumulative 1 AU crossings in a heliocentric coordinate system corotating with the Sun, for protons 80--1300 MeV, for the GLE 71 event. The center of the injection region is at N11W76 in this plot. The mean free path is $\lambda$=0.3 AU.}
   \label{fig.map_gle71} 
     \end{figure}
%

%
%-------------------------------------- Two column figure - comparison with PAMELA data - intensities
   \begin{figure*}
   \centering
   \includegraphics[scale=.48]{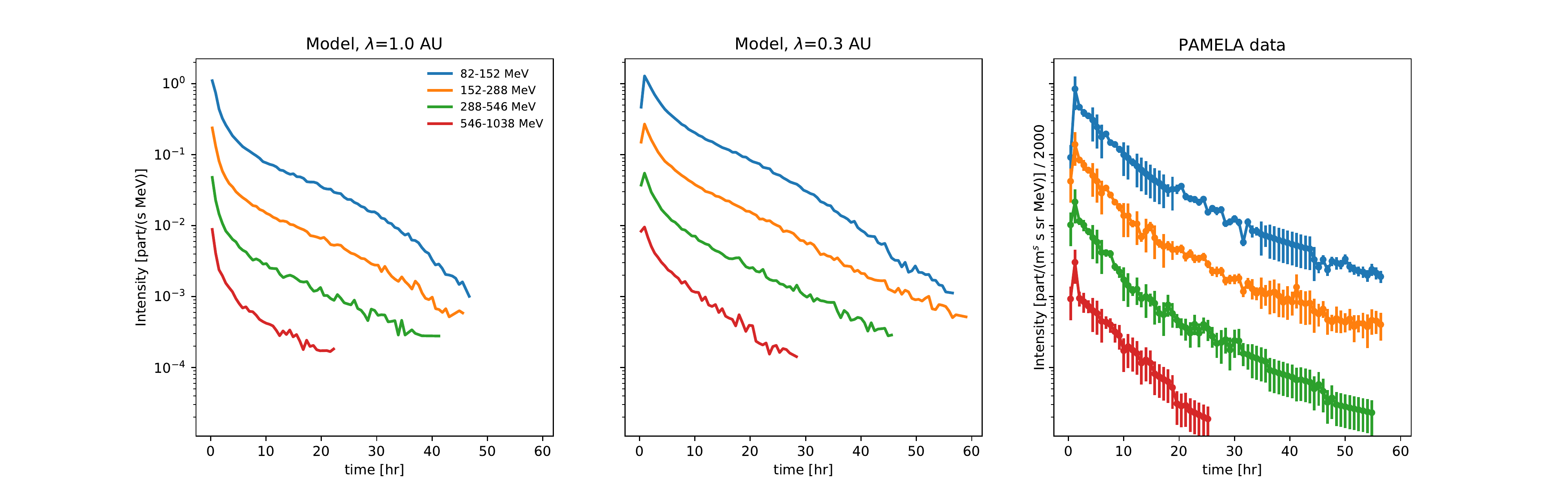}
   \caption{Time-intensity profiles of protons in four energy ranges, for the GLE 71 event. The left and middle panels show simulation results for an observer at the location of Earth, for  $\lambda$=1.0 AU and 0.3 AU respectively. The right panel shows preliminary data from PAMELA.}
   \label{fig.pamela_comp} 
     \end{figure*}

In addition to performing idealised runs, as described in Sections \ref{sec.monoe} and \ref{sec.power}, we applied our 3D relativistic proton simulations to a specific SEP event for which PAMELA data were made available to us by the PAMELA collaboration. The event is GLE 71, occurring on May 17th, 2012. 
Here we present the results of initial simulations in which we considered protons in the energy range 80--1300 MeV, injected instantaneously with a power law spectrum with $\gamma$=2.8 from a region at 2 solar radii, centered on the flare location, which was N11W76. The final time of the simulation was 61 hours. A solar wind speed  $v_{sw}$=400 km s$^{-1}$ was used. For most GLE events,  detailed information on the extent of the injection of relativistic protons near the Sun is not available, but there are indications that it is much smaller compared to lower energy protons \citep{Gop2012}, i.e. that only the nose of the shock is an efficient accelerator at the high energies. Therefore we chose an injection region of 40$\times$40$^{\circ}$ (with the size being the same at all energies considered here) and assume a constant acceleration efficiency within it. The number of particles in the simulations was $N$=3,000,000.

GLE 71 was studied in detail in an earlier publication which focussed on comparing simulations with multi-spacecraft SEP data for energies up to $\sim$100 MeV  \citep{Bat2018b}. In that work, the tilt angle best fitting the conditions during the event was found to be $\alpha_{nl}$=57$^{\circ}$, within an $A^-$ IMF configuration (see Figure 3 of  \cite{Bat2018b}). The HCS for this event is more \lq wavy\rq\ than the one seen in Figure \ref{fig.map_plots}. 

We carried out simulations for two values of the mean free path, assumed to be independent of energy,  $\lambda$=1.0 AU and 0.3 AU.
Figure \ref{fig.map_gle71} shows a map of 1 AU crossings for the simulation with  $\lambda$=0.3 AU.  Because the extended injection region is wider than in the simulations presented in Section \ref{sec.maps} and it intersects the HCS, a strong HCS drift (eastward because of the $A^-$ IMF configuration) is observed.

The source region for GLE 71 was magnetically well connected to Earth, so that an Earth observer was located at a position $[\Delta\phi_{1AU}, \Delta\delta_{1AU}]$=[-2$^{\circ}$, -13$^{\circ}$] with respect to the centre of the injection region at the time of the flare., i.e. within the 40$\times$40$^{\circ}$ injection region. Therefore drift along the HCS did not play a role in determining SEP arrival in this event.
To derive simulated intensity profiles near Earth we collected counts within a 10$\times$10$^{\circ}$ collection tile to ensure good statistics.  

Figure \ref{fig.pamela_comp} shows the intensity profiles in four energy channels for simulations with mean free path $\lambda$=1.0 AU ({\it left}) and 0.3 AU ({\it middle}), compared with the PAMELA intensity profiles ({\it right}).
The PAMELA data shown are based on omni-directional measurements taken in low Earth orbit. However, they account for a correction factor related to the pitch angle anisotropy registered during the first few hours, in particular the first polar pass (see \citealt{Adr2015}). The PAMELA intensity time profiles are broadly consistent with those measured by the Energetic Proton, Electron, and Alpha Detector (EPEAD) and High Energy Proton and Alpha Detector (HEPAD) instruments on board the Geostationary Operational Environmental Satellites (GOES) 13 and 15.
  
It is useful to comment on the qualitative differences between the two simulations and on the comparison with the data.
Regarding the initial phase of the event, it is noticeable that the peak intensity is reached very quickly in the $\lambda$=1 AU simulation, while in the $\lambda$=0.3 AU case peak intensities are reached after a longer time, in a way that matches the observations better. Following the peak, intensities decay rapidly for $\lambda$=1 AU, another feature that is not present in the data, suggesting a situation with stronger scattering, better reproduced by the simulation with $\lambda$=0.3 AU. Following the initial fast decay, the slope of the intensity time profile increases in the simulations, to values closer to those of the observations.

Starting around t=30 hrs, in the $\lambda$=1 AU modelled intensities, an increase in the slope of the decay is seen  in the lowest energy channels, which is not present in the PAMELA data. The same effect is visible, though to a lesser degree, in the  $\lambda$=0.3 AU simulation. This behaviour results from loss of magnetic connection of the observer to the flux tubes within which acceleration took place (the injection region in our simulation), as a result of corotation.
In the higher energy channels the profiles are smoother and the sharp change in the decay time constant t=30 hrs is not observed, as a result of drift effects being stronger, with more efficient leaking of particles out of the injection flux tubes.
It should be noted that turbulence-associated perpendicular transport, not included in our simulations, would smooth the difference in the intensity between the flux tubes connected to the acceleration region and those not connected.

Overall, the simulation with  $\lambda$=0.3 AU appears to produce a better fit in the lower energy channels, although at the higher energies the simulated decay is faster than in the PAMELA data. This may indicate that the size of the injection at higher energies is smaller than for lower energies, or may be related to an energy dependence of the mean free path. Both effects are not included in the initial simulations presented here. 
It should be noted that although the values of mean free path are very different, the slope of decay is not dissimilar in the two simulations of Figure \ref{fig.pamela_comp} . This is unlike the behaviour in 1D simulations, in which the slope is a strong function of the mean free path.

\section{Discussion and conclusions} \label{sec.discuss}

We presented 3D simulations of relativistic proton propagation from the Sun to 1 AU, which included 3D effects associated with particle drift and the presence of a HCS. We considered monoenergetic and power-law populations, injected from a small region at the Sun, to study the qualitative aspects of 3D propagation.
In addition we performed initial simulations for an extended injection region aiming to reproduce PAMELA observations for the GLE 71 event. In further work, we plan to extend the modelling to other PAMELA events.

It should be stressed that our simulations have focussed on the role of IMF polarity and HCS, while other potentially important processes such as perpendicular scattering and magnetic field line meandering \citep{Lai2016} have not been included. The injection of relativistic protons has been assumed to be instantaneous and located near the Sun, which is a reasonable approximation at these energies \citep{Gop2012}.
  In recently published work, \cite{Koch2020} presented an analysis of the 2017 September 10 GLE event using a 2D model of particle transport that includes perpendicular diffusion due to magnetic field turbulence. We expect that inclusion of such effects within our 3D model would smooth particle intensity profiles, eliminating discontinuities at low energies that are due to loss of connection to the flux tubes within which acceleration took place, while retaining the important effects of IMF polarity and HCS. We plan to carry out this study in future work.

The main conclusions from our analysis are as follows:
\begin{itemize}
\item
Propagation of relativistic protons is strongly influenced by the IMF polarity (via gradient and curvature drifts) and the HCS (via HCS drift), making a 3D description necessary. Relativistic protons are not confined to the magnetic flux tube in which they were injected. They experience dilution within the interplanetary medium much faster than $\sim$10 MeV protons, making their detection at a given location less likely than at lower energies. Corotation is less important at relativistic energies compared to lower proton energies.
In contrast to the 1D approach, leakage from the magnetic flux tube in which the particles were injected is a key new phenomenon within a 3D description.
\item 
There are significant differences in the relativistic proton propagation for $A^+$ and $A^-$ configurations, a fact not captured by 1D models, which do not include IMF polarity and a HCS. The average number of 1 AU crossings is significantly larger for $A^-$ than for  $A^+$, due to efficient outward HCS drift in the latter configuration. Maps of 1 AU crossings show that $A^+$ configurations are characterised by stronger HCS-induced propagation across heliolongitudes compared to $A^-$.
\item
  In 3D, injection properties of the SEP population are processed by transport, making intensity profiles and spectra strongly observer dependent. Fluence spectra at 1 AU do not reflect the injection spectra. The decay constant of intensity profiles is not related to the mean free path in a simple way, as is the case in 1D models.
  \item
Comparison of our simulation results with PAMELA observations in the energy range 80 MeV -- 1 GeV for GLE 71 (May 17th 2012) shows that resonable agreement with data can be obtained by choosing an injection region of 40$\times$40$^{\circ}$, a mean free path $\lambda$=0.3 AU and injection spectrum with $\gamma$=2.8. Varying any of these parameters, as well as modifying assumptions on the energy dependence of $\lambda$ and injection region size will influence the intensity profiles. Such an analysis will be the subject of future study.
\end{itemize}

\begin{acknowledgements}
This work has received funding from the UK Science and Technology Facilities Council (STFC) (grant ST/M00760X/1) and the Leverhulme Trust (grant RPG-2015-094). We thank the PAMELA Collaboration for providing us with preliminary results concerning the 2012 May 17 event. G.A.$\,$de$\,$N. acknowledges  support from the NASA/HSR grant NNH13ZDA001N-HSR, and the NASA/ISFM grant HISFM18. The work of J.G. was supported by NSF under grant 1931252.
\end{acknowledgements}

\bibliographystyle{aa} % style aa.bst
\bibliography{highen_biblio} 

\clearpage

\end{document}